# A deep learning approach for direction of arrival estimation using automotive-grade ultrasonic sensors

Mohamed Shawki Elamir
TU-Ilmenau
Valeo Schalter und Sensoren
mohamed-elamir.mohamed@tu-ilmenau.de

Heinrich Gotzig
Valeo Schalter und Sensoren
heinrich.gotzig@valeo.com

Raoul Zöllner
HS-Heilbronn
Raoul.zoellner@hs-heilbronn.de

Patrick Mäder
TU-Ilmenau
patrick.maeder@tu-ilmenau.de

## ABSTRACT

In this paper, a deep learning approach is presented for direction of arrival estimation using automotive-grade ultrasonic sensors which are used for driving assistance systems such as automatic parking. A study and implementation of the state of the art deterministic direction of arrival estimation algorithms is used as a benchmark for the performance of the proposed approach. Analysis of the performance of the proposed algorithms against the existing algorithms is carried out over simulation data as well as data from a measurement campaign done using automotive-grade ultrasonic sensors. Both sets of results clearly show the superiority of the proposed approach under realistic conditions such as noise from the environment as well as eventual errors in measurements. It is demonstrated as well how the proposed approach can overcome some of the known limitations of the existing algorithms such as precision dilution of triangulation and aliasing.

## 1. INTRODUCTION

The topic of autonomous driving is gaining more and more attention from researchers and from the automotive industry leaders, especially over the past few years. One critical element of autonomous driving is environmental perception. In a few words, environmental perception means the awareness of the surrounding environment of the vehicle in terms of existing obstacles in the vicinity of the vehicle and its path as well as the environmental conditions such as the temperature and the humidity levels. This information is not only useful for the goal of achieving autonomous driving, but also for advanced driving assistance systems (ADAS) that aid the driver and are considered as building blocks and early stages of fully autonomous driving and higher functionalities such as automatic parking. Environmental perception is carried out using sensors. There are several types of sensors that are used in this domain such as radars, lidars, cameras and ultrasonic sensors. The quality of the information provided by these sensors in terms of precision and accuracy directly influences the ability of the ADAS to correctly perform its functionality. In this paper the focus is on one of these sensor types which is the ultrasonic sensor, and the target is to estimate the direction of obstacles off which the ultrasonic waves are reflected under realistic conditions as described in Figure 1-1. The motivation and target of this work is to obtain higher quality information from the ultrasonic sensors using the proposed deep learning-based approach for processing the data output from these sensors.

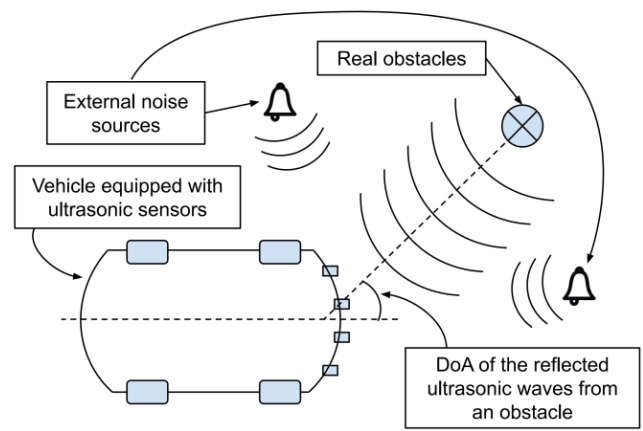

**Figure 1-1: Using an array of ultrasonic sensors for the direction of arrival estimation of an incident echo reflecting off an obstacle in the vicinity of the ego vehicle**

Ultrasonic sensors are used in the automotive industry specifically for near obstacle detection. Ultrasonic sensors belong to the category of short-range distance sensors. They function up to a range of approximately 7 meters. The sensors report the distance between the ego vehicle and an obstacle in the vicinity of a vehicle through echolocation techniques and Euclidean triangulation. In principle, the ultrasonic sensor fires an ultrasonic wave at a certain time $t_{send}$. This wave propagates though air till it hits an obstacle in the immediate environment of the vehicle and an echo is reflected from this obstacle. The ultrasonic sensor detects the reflected echo at time $t_{receive}$. By knowing the speed of propagation of ultrasonic waves in air and the time taken by the fired wave to propagate back and forth between the sensor and the obstacle, the radial distance between the obstacle and sensor can be directly calculated. A vehicle is usually equipped with several ultrasonic sensors in one bumper. By calculating the radial distance between the obstacle and several sensors in the bumper, the position of the obstacle can be determined in a 2D map around the planner view of the vehicle using triangulation techniques which will be detailed later.

One of the challenges that face this type of obstacle position detection using ultrasonic sensor is that the characteristics of the echo differ based on the shape and orientation of the obstacle. One of these characteristics is the echo duration which typically ranges from a value of a couple of hundred microseconds for a narrow obstacle such as a standard 75 mm thick tube, up to a few milliseconds for a wider obstacle such as a wall. This variation in the duration of the echo poses a degree of uncertainty in the measurement. The problem is which point along the whole echo duration should be taken as the representative for the arrival time of the echo. This problem is especially evident when a modulation technique is used in the fired wave such as binary phase shift keying (BPSK) or chirp where the signal needs to be demodulated upon reception. These few centimeters of uncertainty lead to larger errors in the obstacle position estimation due to the phenomena of triangulation precision dilution. This phenomenon is discussed in more detail later.

On top, there are several factors that could lead to potential errors in the ultrasonic sensor measurement. An example of these factors is the variation in the external temperature. This is problematic due to the fact that the speed of propagation of the ultrasonic wave in air changes based on the temperature of the medium. The speed of propagation is inversely proportional to the air temperature. This problem cannot be simply mitigated using a temperature sensor because of the non-uniformity of the temperature distribution over all the ultrasonic sensors in the vehicle and also along the path of propagation of the wave itself between the vehicle and the obstacle. Another potential source of error is the drift in the reverberation frequency of the sensor membrane due to the aging of the sensor or manufacturing imprecision. These small deviations lead to larger errors, as mentioned earlier, due to triangulation precision dilution.

One solution to this problem is to exploit the phase information for the incident waves and use it to estimate the direction of arrival (DoA) of the echo and identify the exact angle of incidence of the wave front and thus eliminate the uncertainty in the transverse dimension in the 2D planner map view of the vehicle. Combining this approach with the classical triangulation approach will lead to a more precise measurement and thus better and more reliable ADAS functionality. There exist several deterministic algorithms in literature for DoA estimation such as correlation based, ESPRIT and MUSIC and several more. An example of the MUSIC DoA estimation algorithm output is represented in Figure 1-2. There are 2 general drawbacks of the existing DoA estimation algorithms. The first problem is that they are sensitive to errors in measurements. This eventually leads to larger error in the DoA estimation which defeats the purpose for which this approach is used. This error could come from the sensor itself due to aging or could come from external environmental influences such as fluctuating temperature and humidity values or external sources of noise.

The second problem is aliasing which appears if the separation between the sensors is larger than half the wavelength. In this case the range of operation of the DoA estimation algorithms could be limited to only a small range around the axis of the sensor array, or else, the DoAs reported by the deterministic algorithms are a set of possible directions with no predefined way of how to prioritize between them. For benchmarking we select the MUSIC algorithm since on the one hand its performance is converging to the Cramer-Rao bound which is the asymptotic optimal performance and on the other hand it is not computationally expensive and thus would be fair to include it as a realistic means of DoA estimation in an embedded system equipped in a vehicle.

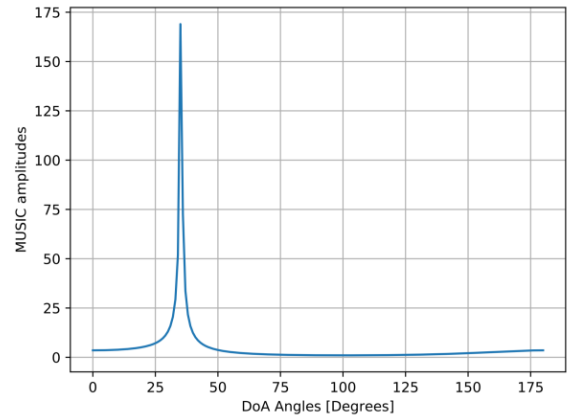

**Figure 1-2: An example of the MUSIC algorithm detecting the DoA of a single source at an angle of 30° on an array of 2 ultrasonic sensors spaced at 0.5 × wavelength, with no noise in the environment**

In this paper we present a deep learning-based approach for DoA estimation that surpasses the optimal MUSIC algorithm under realistic conditions such as the presence of external noise sources and errors in the sensor measurements. This approach is used as an end-to-end solution for estimating with high precision the direction of the object in the vicinity of the vehicles which implicitly combines together the information used in triangulation and the deterministic phase dependent DoA estimation, overcoming the drawbacks of both solutions namely being the triangulation dilution and the aliasing as the distance between the sensor elements is larger than half the wavelength.

## 2. State of the art

There exists a large number of publications in the field of artificial intelligence and deep learning targeting the goal of autonomous driving and tackling the related challenges. For the scope of this paper, we address the category of these publications that are aimed towards utilizing the data provided by typical sensors equipped in vehicles for the perception of the environment surrounding the vehicle. This provides essential knowledge for control systems of a vehicle to navigate in an autonomous fashion. We see publications dating as far back as 1989 such as [1] where a camera is used to provide input for a neural network-based system that performs end-to-end autonomous driving. In recent publications as well, we see work by researchers in Valeo [2] where a network of fish-eye cameras is used to provide a large dataset for training neural network-based systems for semantic segmentation of the vehicle surroundings into predefined classes. Several publications combine different technologies together with the images captured by cameras such as [3] where a methodology is proposed for fusion of information from LIDAR and thermal camera images to perform autonomous driving. Other perception sensors are also employed for identifying obstacles without relying on visual information from cameras such as the very recent publication [4] where the LIDAR information on its own is exploited through a deep learning structure for the identification of obstacles through a generative-adversarial approach. Similar publications employing deep learning approaches for extraction of information of interest from raw data from other RADARs for the purpose of autonomous driving is presented in recent publications such as [5], [6] and [7], where in the latter we see on top a fusion approach for both the RADAR and LIDAR data. Ultrasonic sensors, as well, are used for environment perception using deep learning based methodologies, such as in [8] where deep learning approaches are used to classify noise and echoes of automotive-grade ultrasonic sensors and [9] where a deep learning structure is proposed to suppress spurious noise artifacts superimposed on ultrasonic echo raw signals showing that it surpasses conventional methods in terms of maintaining the integrity of the signal of interest and causing minimal distortion.

Environmental perception in ultrasonic sensor-based systems is accomplished through triangulation of the distances reported by individual sensors as is detailed later in this work. The main drawback facing this localization method is the triangulation dilution problem. This problem is well presented in [10] and [11] where it is presented for robot environment perception algorithms and self-localization methods. A few proposals of how localization problems could be mitigated is presented in [12] where 3 methods are presented to overcome several localization issues such as blind spots and beacon order, but still it is limited by the dilution limitation. In this work we show how the proposed deep learning technique combines the direction of arrival estimation information with the individual sensor distance information to overcome the triangulation dilution problem and provide more accurate estimation of the obstacle position in a planner 2D vehicle surrounding map.

There are several approaches used to identify the source DoA of a signal's source by using a sensor array and analyzing the phase shift information of the received beacon on each element. In these papers [13], [14] surveys of the existing DoA estimation algorithms are presented showing how the path delays estimation, and accurate channel estimation between a transmitter and an array of receivers can be used to identify an incident wave's DoA. A group of these methods has shown excellent performance that is approaching the Cramer Rao asymptotical theoretical best performance. This is called the parametric DoA estimation algorithms, which includes the MUSIC algorithm [15] that was first introduced in 1986 and the ESPRIT algorithm [16] that was proposed a few years later in 1989. A very good comparison and analysis of these 2 methods is presented in this paper [17]. In this work the MUSIC algorithm is used as benchmark since it is one of the most widely used DoA estimation algorithms and it is approaching the theoretical optimum performance. Further on, results are present showing how a deep learning approach combining several input information can surpass the performance of MUSIC algorithm under realistic conditions of environment noise and sensor measurements errors.

In recent literature, the topic of using deep learning for DoA estimation is gaining more and more interest. In very recent publications in 2020 [18] A feed forward neural network approach (FNN) for DoA estimation is presented. Results are benchmarked against the MUSIC algorithm showing how performance could be surpassed in case of having multiple unknown number of signal sources. Another publication [19] presents the superiority of deep learning approach compared to eigen decomposition DoA estimation algorithms such as MUSIC algorithm from a computational complexity point of view. In this paper [20], strong results are presented showing how a deep learning-based approach for DoA estimation surpasses the MUSIC algorithms given imperfections in the receiving sensor array since the deep learning approach is data driven and compensates for these imperfections through the learning process. This work differs from the existing publication in the fact that the focus is on how a deep learning approach surpasses the conventional DoA estimation algorithms given the realistic condition of having imperfections present in the received signal itself and how it is corrupted by external environment factors such as noise.

## 3. Deterministic algorithms

The function of ultrasonic sensors in a vehicle is to perform echo location. This includes detecting the presence of an obstacle in the vicinity of a vehicle and reporting the radial distance between the sensor and this obstacle. As mentioned, an ultrasonic based system is composed of several sensors. The system collects the radial distances reported by the individual sensors and calculates the position of the obstacle with respect to the vehicle in a 2D planner view map of the vehicle and its immediate surrounding environment. In the following subsections we will present how echo location is performed through triangulation and the challenges facing triangulation, namely precision dilution. We will also present how this could be mitigated through a DoA estimation algorithm. The different types of DoA estimation algorithms will be presented with a more detailed focus on the MUSIC algorithm. The drawbacks of the DoA estimation algorithms will be highlighted as well leading to the proposed deep learning-based approach that addresses these drawbacks and provides a superior performance.

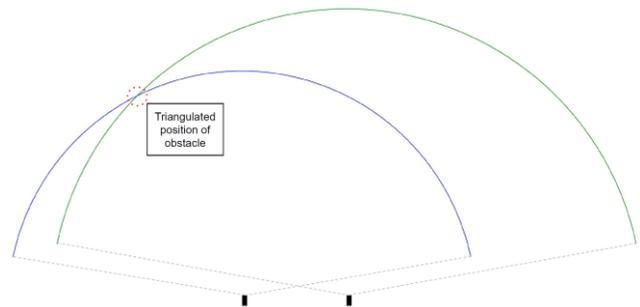

**Figure 3-1: A simple triangulation example where a potential obstacle exists at the intersection point of the radial distances reported by 2 ultrasonic sensors**

## 3.1 Triangulation

### 3.1.1 Concept

The radial distance between the individual ultrasonic sensors and an obstacle is reported to the system controlling these sensors. At the center of this system is usually a microcontroller in which the triangulation calculations are performed. Each 2 neighboring sensors reporting distances create a point of intersection in the systems 2D planner map as depicted in Figure 3-1. The Cartesian coordinate position of this intersection with respect to the origin of the vehicle is stored as a possible obstacle. Usually, the ultrasonic based systems report the presence of an obstacle after several confirmations to increase the robustness against environmental noise sources and to prevent the false indication of non-existing obstacles.

If more than 2 neighboring sensors detect the presence of an obstacle and report the corresponding distance, then the system will have an over determined set of non-linear equations. Best case scenario would be if all the 3 distances intersect in only one point. In this case the solution would be trivial since one of the three measurements is redundant and conveys no extra information. In case the distances do not intersect in a single point, which is the more probable and realistic case, there are specific algorithms such as Gauss-Newton method which relies on the calculation of the Jacobian matrix based on the measurement models and going through iterations to minimize the error in solving this over determined set of non-linear equations. There is also the Nadler-Mead method which is a numerical analysis method which also relies on iterations until the optimal solution is reached. This type of approach is usually very computationally expensive and not suitable for implementation in an automotive-grade industrial microcontroller, especially with the requirements of real-time processing of the measurements unless some heavy optimization and pre-assumptions are made. This is not the focus of this work since the sensor array at hand is composed of 2 elements and thus the focus will be only on the first and simpler scenario with the intersection of only 2 radial measurements.

### 3.1.2 Drawbacks

Precision dilution of triangulation is an artifact of geometry which is caused by the limited space on the bumper of a vehicle causing the sensors to be packed closer together typically with a distance not larger than 50 cm between each 2 neighboring sensors.

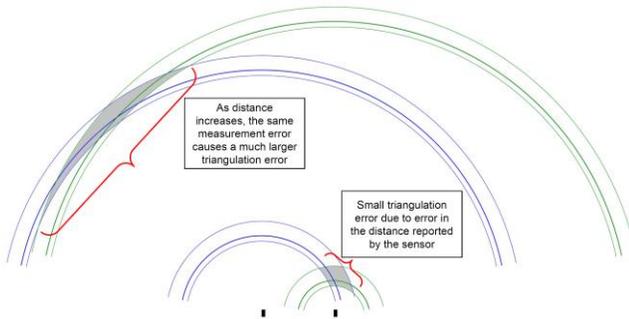

**Figure 3-2: Excessive increase in localization error with the increase in distance between the potential obstacle and the ultrasonic sensors given a constant level of measurement error**

This causes an increasing error in the localization precision of the intersection point between the distances reported by 2 ultrasonic sensors as the line-of-sight (LOS) of each of the sensors to the obstacle becomes more and more parallel to the other. In Figure 3-2 we see the gray area which is the area of potential error in localization due to the same value of error standard deviation in the sensor but at different distances. We see that the potential localization error increases to larger levels as the distance from the sensor increases.

A common method to overcome the physical limitation of precision dilution is to employ a DoA estimation algorithm to estimate the angle of the incident wave. The intersection between the triangulated area of spatial uncertainty, in grey, and the line originating from the sensor with an angle equivalent to the DoA of the incident wave, indicates the most probable location of the obstacle and thus eliminating, or at least reducing, the spatial uncertainty. In the next section we discuss the chosen DoA estimation method and its limitation and afterwards the proposed approach is presented.

## 3.2 DoA estimation

There are different types of DoA estimation algorithms. Here we choose the MUSIC algorithm to be the benchmark because its performance is proven to be optimum and converging towards the Cramer Rao bound. Another reason for choosing the MUSIC algorithm is that it is computationally acceptable in terms of complexity and thus could be pragmatically used in embedded systems for the sensors equipped in vehicles.

### 3.2.1 Concept

For the MUSIC algorithm the covariance matrix is constructed from the received signal. Then, an array of steering vectors is synthesized. They contain all possible incident wave directions on the sensor array, with a predetermined angular resolution. Each steering vector is projected onto the null space of the covariance matrix. The result is that the smallest value possible occurs when the steering vector corresponds to the DoA of one of the signals contributing in the creation of the covariance matrix. Thus, a peak is attained by inverting the resulting value. This is achieved by raising the result to the power of -1. A few simplifications are also applied to reduce the computational complexity without compromising the mathematical integrity of the results.

The MUSIC algorithm estimates the DoA of a signal following the data model described in the following equation.

$$Y = A(\theta)S + N_o$$

With $Y$ having a dimension of $M \times 1$ as the output of the sensor array. $A$ is the $M \times N$ signal subspace where $M$ is the number of elements of the receiving sensor array and $N$ is the size of the signal. $\theta$ is the angle of the DoA of the signal, $S$ having a size $N \times 1$ is the receiving signal coefficients and $N_o$, also of size $M \times 1$, is assumed to be additive white Gaussian noise (AWGN).

The covariance matrix of the output of the array is described in the following equation.

$$R_y = A(\theta)R_s A(\theta)^H + \sigma^2 I$$

Where $R_s$ is the covariance matrix of the signal coefficients vector and its size is $N \times N$. $\sigma$ is the standard deviation of the Gaussian noise. $R_y$ is the covariance matrix of the received signal and is of size $M \times M$.

Eigen vector decomposition is applied to the covariance matrix of the output of the sensor array to be able to get the basis vectors of the null subspaces of the sensor array output vector. This step generates $N$ eigen values $\lambda_i$ and correspond to $N$ eigen vectors $u_i$. The null vectors are grouped in a matrix and the sensor array signal output $I_s$ is projected into the null subspace spanned by the null vectors matrix. This expression is then reduced without losing generality and inverted leading to the final expression of the MUSIC algorithm which is represented in the following equation.

$$P(\theta) = \frac{a^H(\theta)a(\theta)}{a^H(\theta)V_n^H a(\theta)}$$

Where $a(\theta)$ is the standard unitary excitation of the sensor array for an incident signal from a direction having an angle $\theta$, thus having the dimensions of $M \times 1$. $V$ is the null space basis matrix of dimensions $M \times M$, built using the null space basis vectors composing matrix of dimensions $M \times (M - D)$, where $D$ is the number of expected incident signals.

In this work we are using a sensor array of 2 elements which allows for the DoA estimation of a single signal. In this case $M$ will be set to 2 and $D$ is always 1 leading to the reduction of the null space basis matrix to a vector of dimensions $M \times 1$.

### 3.2.2 Drawbacks

The correctness of the result is highly dependent on the exactness of the measured data from the sensor. Errors in measurement will directly lead to exceeding errors in the results of the MUSIC algorithm.

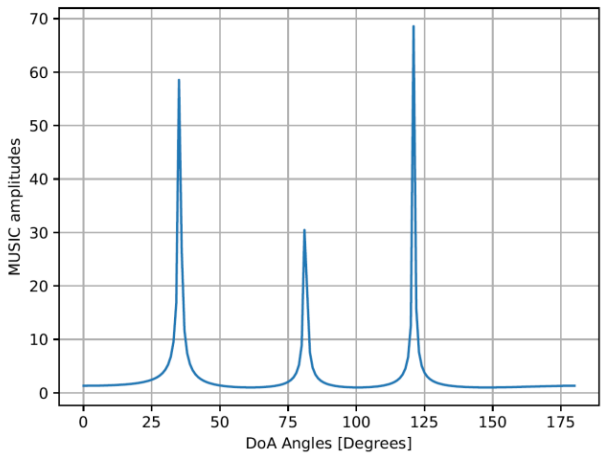

**Figure 3-3: Aliasing problem seen when 2 sensors are positioned with separation of $1.5 \times$ wavelength, and an incident wave is at an angle of $30°$**

Another drawback is the aliasing that occurs when the distance between the sensors is larger than half the wavelength. The ultrasonic sensors driver assistance systems function in the range between 40 and 65 kHz. Therefore, the wavelength of the signal is in the range of 0.2 – 0.4 cm. The diameter of the membrane casing is usually in the range of double this value due to series production constraints. This leads to the fact that the sensor elements will never be placed at a distance where aliasing could be prevented. An example of aliasing when the sensors are placed at $1.5 \times$ wavelength distance from each other and an incident wave is at an angle of $30°$ is represented in the Figure 3-3.

In the next section, a deep learning-based approach is proposed which addresses both these points and surpasses the MUSIC algorithm performance under more realistic conditions where the environment factors such as noise and measurement errors are taken into consideration.

## 4. Proposed deep learning approach

As detailed earlier, the classical approach of triangulating obstacles in the vicinity of a vehicle has drawbacks due to the phenomena of triangulation precision dilution. DoA estimation algorithms could be used such as MUSIC to exploit the phase information and estimate the direction of arrival of the signal and thus giving more precision to the localization problem. This approach also suffers from drawbacks due to sensitivity of this family of algorithms to any drift in the measured values or influence from additive environment noise. This could be due to changes in environmental conditions or due to errors in the measurement itself. Another problem is the aliasing problem when the spacing between the sensor elements is more than half the wavelength of the ultrasonic wave. In this case the classical algorithms report more than one possible DoA without being able to prioritize between them. In this work, a deep learning-based approach is presented for direction arrival estimation to overcome these challenges.

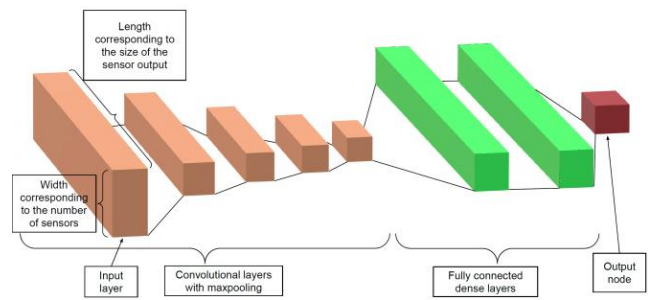

**Figure 4-1: Neural network-based deep learning structure exploiting the convolutional and the dense layered architecture**

A convolutional neural network (CNN) is used with 5 convolutional layers having 64 2-dimensional kernels of size $16 \times 4$. The digitized signal from the ultrasonic sensor is a one-dimensional signal. Since we are assessing the signal coming from 2 sensors therefore the kernel is designed to take in information from both sensors simultaneously and thus the kernel size is 2-dimensional with the value of the added dimension corresponding to the double number of sensors in the array. The double of the number of sensors is needed because for each sensor there exists the real and imaginary parts of the signal. This number could then be scaled with increasing the number of sensors in the array. After the first layer, the size of the processed data is compressed through a maxpooling layer of $2 \times 2$ after each convolutional layer. Starting from the third layer the compression is then done by a maxpooling layer of 2 by 1. The activation function used is the rectified linear unit (ReLU). The optimization algorithm used is the ADAM algorithm to avoid getting stuck in a local minimum and the cost function is the mean square error (MSE).

The convolutional layers are followed by 2 dense layers that are fully connected. The output layer is composed of one neuron having the hyperbolic tangent as an activation function. Output is designed in this way to limit the possible values of the output to a range [-1, 1] which is then scaled to the labels of the recorded DoA to correspond to [-pi/2, pi/2] where angle zero corresponds to the axis of the sensor or in other words perpendicular to the membrane of the sensor. In Figure 4-1 a schematic of the employed neural network is presented.

## 5. Testing and validation

Automotive-grade ultrasonic sensors rely on the concept of firing a wave and analyzing the echo reflected from obstacles in the vicinity of the vehicle. By calculating the time of flight between the firing of the wave and the reception of the echo, the radial distance to the obstacle can be estimated. The membrane of the ultrasonic sensor is driven to reverberate for a few cycles and then is made to stop and be in the listen mode waiting for the echo reflection from the surrounding environment. A simulation environment is constructed with these conditions for testing, validation and software prototyping. A signal capturing device is constructed as well for the purpose of fast validation of the results. Both simulation environment and the data capturing setup are presented in this section.

### 5.1 Simulation setup

In the simulation environment, the reverberation frequency of the sensor is set to the nominal value of 51.2 kHz. And the speed of propagation of the ultrasonic wave in air is set to a nominal value of 340 m/s. The sampling rate is set to 1 (MSPS). The echo envelope shape is based on the shape of the echo reflected from a standard test obstacle which is a 1 m high tube with a cross section diameter of 75 mm. Additive white Gaussian noise (AWGN) is added with different signal to noise ratios (SNRs) to evaluate the performance of the legacy algorithms as well as the proposed approach under different levels of disturbance noise.

A simulated signal of two sensors placed at an optimal distance from each other is recorded for different SNR values. The optimal distance is equivalent to half the wavelength of the simulated ultrasonic wave, thus allowing for a full range of [-90, 90] degrees of DoA estimation. The aperture angle of a typical automotive-grade sensor is in the range of 90 to 120 degrees forming a conical shape around the axis of the sensor. For the simulation purposes the maximum range is selected. An example of the simulated signal without noise is presented in Figure 5-1 and another example with SNR of 0 dB is in Figure 5-2. It is clear how the AWGN distorts the shape of the signal and thus hinders the ability of MUSIC algorithm to perform optimally.

An SNR range of [-30, 20] dB is simulated and the 2 approaches for DoA estimation are tested on the simulated data. The neural network of the deep learning approach is trained over 80% of the generated data and then tested over the remaining 20%. The performance of the 2 approaches is compared in terms of absolute error. When the MUSIC algorithm fails to converge to a specific angle, then the angle 0 is returned which is corresponding to the direction along the axis of the sensor. The same treatment applies for the deep learning approach.

### 5.2 Data acquisition setup

The signal from the sensor is sampled with an analog to digital converter (ADC) at a rate of 1 MSPS. The data is collected by an ARM M7 micro-controller and then sent via Ethernet to PC for post processing. Due to hardware construction limitations, the sensor elements could not be placed at less than half the wavelength from each other. The closest possible arrangement is at 1.5 × wavelength. This introduces a strong limitation on the detection range of the MUSIC algorithm. We show in the following results how the deep learning-based solution can estimate the direction of the obstacle beyond this limited range.

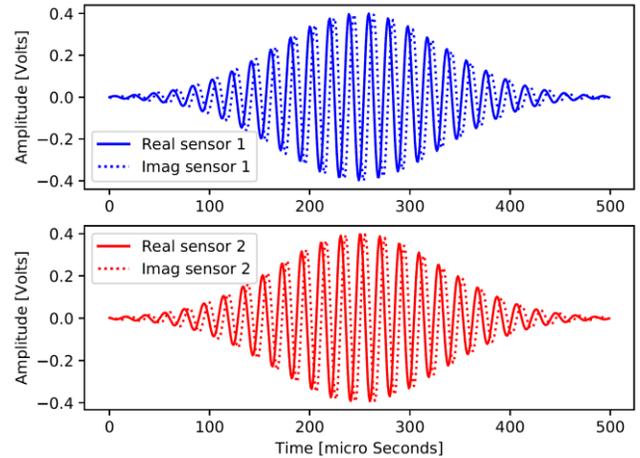

Figure 5-1: An example of the signal simulated on the 2 sensors forming the array with no noise

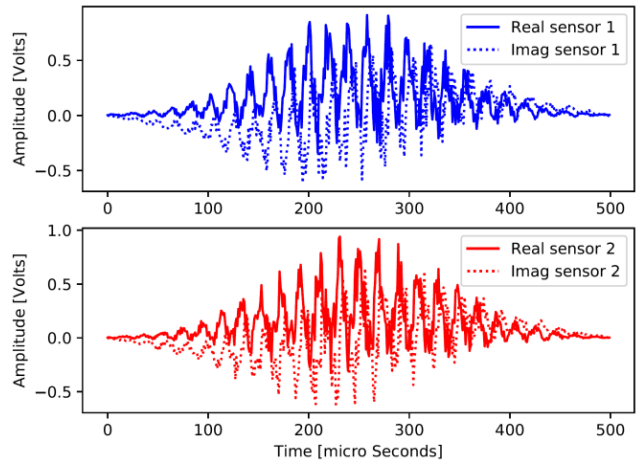

Figure 5-2: An example of the signal simulated on the 2 sensors forming the array with and SNR level of 0 dB which means that the noise energy and the signal energy are equal

This is attributed to the fact that the structure is fed not only with the phase information but with the whole signal. Therefore, the neural network structure can extract spatial distance information to an extent that would allow for prioritization and favoring of one of the possible DoAs. This gives a clear advantage over the deterministic DoA estimation algorithms and allows for overcoming the aliasing limitation as mentioned earlier.

A massive measurement campaign is carried out to record the echo signal reflected from different obstacles such as tubes of different cross sections, walls, boxes, and bushes as an example of a complex obstacle with numerous reflection points. A noise gun is used to fire pre-recorded ultrasonic noise towards the sensors. Different noise sources are used such as truck breaks, a motorcycle, sweeping deep, rain and an air gun. The strength of the noise is controlled to assess the performance of the 2 sets of algorithms under different SNR values.

## 6. Results and discussion

The simulation setup and the data capturing environment are used in the analysis and validation of the existing DoA estimation algorithms with conventional methods as well as with the proposed deep learning approach.

### 6.1 Simulation results

The results comparing the performance of the MUSIC algorithm and the deep learning-based algorithm are represented in Figure 6-1.

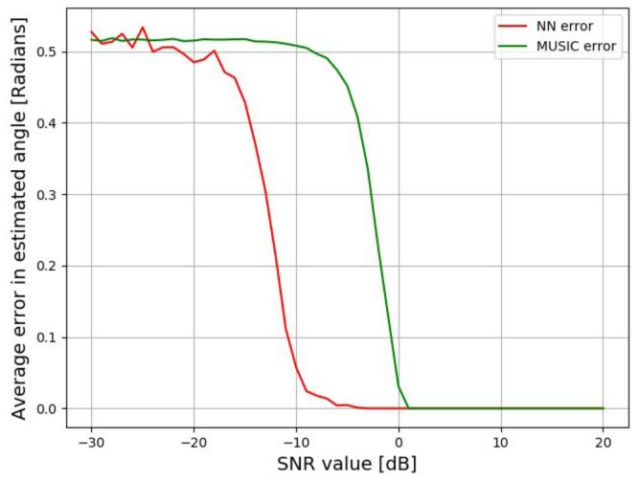

Figure 6-1: The error in the estimated angle by both the MUSIC algorithm and the deep learning-based approach against different levels of SNR values

From these results it is clear that for high values of SNR both the MUSIC and the deep learning approach perform optimally, and their performance is comparable in terms of absolute error in the estimation of the DoA of the simulated signal. When SNR value decreases below 0 dB the deep learning approach surpasses the MUSIC algorithm in performance. As the SNR value keeps dropping the MUSIC algorithm fails completely to function while the deep learning approach still functions with acceptable performance. The results indicate roughly a 10 dB advantage for the deep learning approach.

### 6.2 Measurement results

The results from the measurement campaign is divided into 2 parts. In the first part, the angular range around the axis of the sensor membrane is limited to the range where there is no risk of aliasing and thus both the deep learning-based approach as well as the MUSIC algorithm can function properly. The result of using these 2 approaches is presented in Figure 6-2. The second part of the results is for measurements with an angular value beyond the range where MUSIC could function due to aliasing which originates from the physical constraints of not having the 2 sensor elements at a distance less than half the wavelength from each other. These results are, therefore, presented for the deep learning approach only, since the DoAs are beyond the capability range of the MUSIC algorithm. These results are presented in Figure 6-3.

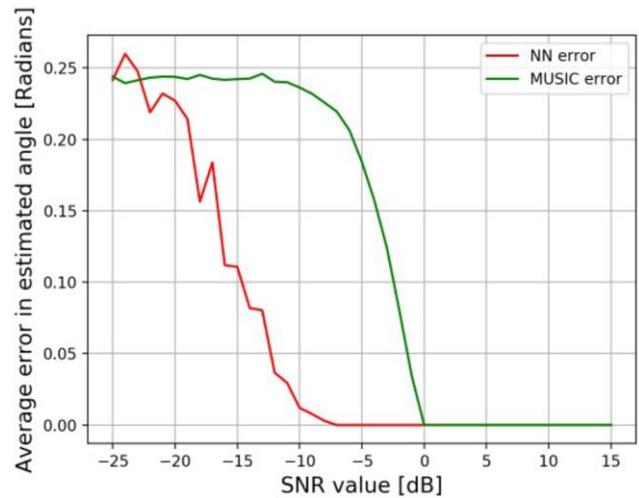

Figure 6-2: The error in the estimated angle by both the MUSIC algorithm and the deep learning-based approach against different levels of SNR values over the limited range of +/- 30° to avoid aliasing

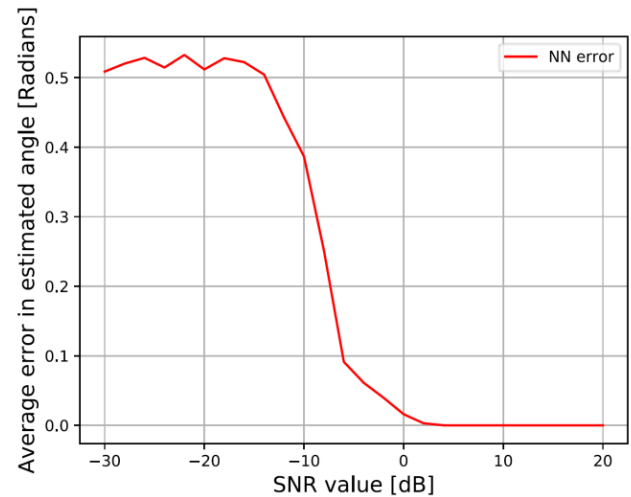

Figure 6-3: The error in the estimated angle by the deep learning-based approach, against different levels of SNR values over the whole angular range, excluding +/- 30°

As we can see from both result sets, the performance of the deep learning approach for DoA estimation surpasses the MUSIC deterministic algorithm by providing better estimations of the DoA under the different ranges of SNR. The improvement matches the simulation results to a large extent showing that the simulation setup is of high fidelity and that the assumed influencing parameters considered, match the reality of the wave propagation characteristics. The performance of the deep learning algorithm maintains the same level of performance outside the limit functional range of the MUSIC algorithm due to the increased spacing between the sensor array elements. This shows how the deep learning approach overcomes the aliasing limitation and provides superior performance for DoA estimation over the complete range of possible angles in the range of the sensor.

## 7. Conclusion

The deep learning approach for DoA estimation is superior to the classical algorithms under realistic conditions such as measurement errors and external noise influences. From the experiments carried out in this work, we see that an improvement of more than 10 dB is achieved using the deep-learning approach. On top, this approach provides means to overcome aliasing since this approach distinctly favors one of the probable aliased DoAs, which is not possible using classical algorithms such as MUSIC. The level of performance in estimating the DoA is maintained over the aliasing angular values.

Combined with the classical triangulation approach, the information attained from the deep learning-based approach from DoA estimation provides a higher level of precision in the localization problem, thus mitigating the precision dilution problem and allowing for higher quality data to be processed by upper layers in the driver assistance systems. This provides for more reliable building blocks for the construction of autonomous vehicles. It also shows that there is great potential for using deep learning-based approaches for signal processing and useful data extraction in comparison to the existing classical deterministic approaches.